\documentclass[twocolumn,aps,prl,showpacs,superscriptaddress]{revtex4-1}
\usepackage[latin9]{inputenc}
\usepackage{amsmath}
\usepackage[dvipdfmx]{graphicx}

\makeatletter

\usepackage{verbatim}
\usepackage{color}
\usepackage{subfigure}
\usepackage{here}

\makeatother
\begin{document}

\title{Half-Integer Shapiro Steps in a Short Ballistic InAs Nanowire Josephson Junction} 

\author{Kento Ueda}
\email{kento.ueda@riken.jp}
\affiliation{Department of Applied Physics, University of Tokyo, 7-3-1 Hongo, Bunkyo-ku, Tokyo 113-8656, Japan}

\author{Sadashige Matsuo}
\email{sadashige.matsuo@riken.jp}
\affiliation{JST, PRESTO, 4-1-8 Honcho, Kawaguchi, Saitama 332-0012, Japan}
\affiliation{Center for Emergent Matter Science, RIKEN, 2-1 Hirosawa, Wako-shi, Saitama 351-0198, Japan}

\author{Hiroshi Kamata}
\affiliation{Laboratoire de Physique de l'\'{E}cole Normale Sup\'{e}rieure, ENS,
PSL Research University, CNRS, Sorbonne Universit\'{e}, Universit\'{e} Paris Diderot,
Sorbonne Paris Cit\'{e}, 24 rue Lhomond, 75231 Paris Cedex 05, France}

\author{Yosuke Sato}
\affiliation{Department of Applied Physics, University of Tokyo, 7-3-1 Hongo, Bunkyo-ku, Tokyo 113-8656, Japan}

\author{Yuusuke Takeshige}
\affiliation{Department of Applied Physics, University of Tokyo, 7-3-1 Hongo, Bunkyo-ku, Tokyo 113-8656, Japan}

\author{Kan Li}
\affiliation{Beijing Key Laboratory of Quantum Devices, Key Laboratory
for the Physics and Chemistry of Nanodevices and Department of Electronics,
Peking University, Beijing 100871, China}

\author{Lars Samuelson}
\affiliation{Division of Solid State Physics
and NanoLund, Lund University, Box 118, SE-221 00 Lund, Sweden}

\author{Hongqi Xu}
\email{hqxu@pku.edu.cn}
\affiliation{Beijing Key Laboratory of Quantum Devices, Key Laboratory
for the Physics and Chemistry of Nanodevices and Department of Electronics,
Peking University, Beijing 100871, China}
\affiliation{Division of Solid State Physics
and NanoLund, Lund University, Box 118, SE-221 00 Lund, Sweden}
\affiliation{Beijing Academy
of Quantum Information Sciences, Beijing 100193, China}

\author{Seigo Tarucha}
\email{tarucha@riken.jp}
\affiliation{Center for Emergent Matter Science, RIKEN, 2-1 Hirosawa, Wako-shi, Saitama 351-0198, Japan}
\begin{abstract}
We report on half-integer Shapiro steps observed in an InAs nanowire Josephson junction.
We observed the Shapiro steps of the short ballistic InAs nanowire Josephson junction and found anomalous half-integer steps in addition to the conventional integer steps.
The half-integer steps disappear as the temperature increases or transmission of the junction decreases.
These experimental results agree closely with numerical calculation of the Shapiro response for the skewed current phase relation in a short ballistic Josephson junction. 
\end{abstract}
\maketitle
\renewcommand{\thefigure}{\arabic{figure}}
\renewcommand{\theequation}{\arabic{equation}}
\setcounter{figure}{0}
The AC Josephson effect has long been studied as a manifestation of macroscopic quantum interference~\cite{joesphson1962possible} as well as for application in the quantum voltage standard~\cite{hamilton2000josephson} and more recently development of superconducting qubits~\cite{KochPRA2007,Devoret2013}.
The Shapiro voltage step, as given by $V = nhf/2e~ (n=1,2,3, ...)$, is an immediate consequence of the AC Josephson effect in the presence of microwave excitation\cite{shapiro1963josephson}, reflecting the phase-mediated binding of the microwave field and the Josephson current.
Therefore, the Shapiro step measurement features the current phase relation (CPR) of the junctions.
The CPR is usually sinusoidal with $2\pi$ periodicity; otherwise, there appear anomalies in the Shapiro steps. 
For example, fractional steps appear when high harmonic components exist in the CPR.
They are theoretically predicted~\cite{Valizadeh2008} and indeed observed in superconductor (SC)-ferromagnet-SC junctions in the vicinity of the $0$-$\pi$ transition~\cite{Sellier2004, Frolov2006, stoutimore2018second}, graphene Josephson junctions~\cite{lee2015ultimately} and PbSnTe Josephson junctions~\cite{snyder2018weak}.
On the other hand, odd-integer multiples of the Shapiro steps are absent in the $4\pi$ periodic CPR, which is observed in topological superconductors hosting Majorana fermions~\cite{rokhinson2012fractional,wiedenmann20164,bocquillon2017gapless,li20184pi}.
The topological features of the SC junctions are attracting intense research interest because of their applicability to topological quantum computation~\cite{nayak2008non,alicea2011non,sarma2015majorana}. 
However, the physics of the bound states are yet elusive.

Josephson junctions of semiconductor nanowires can show both anomalies in the Shapiro responses.
A strong magnetic field invokes the topological transition, resulting in the disappearance of the odd Shapiro steps~\cite{rokhinson2012fractional}.
Furthermore, when the Josephson junctions of the semiconductor nanowires are ballistic but not topological, the CPR is highly skewed~\cite{spanton2017current}; the fractional steps are expected, although they have not yet been observed experimentally.
Studies on the Shapiro steps in ballistic nanowire Josephson junctions can provide important insights on the Andreev bound state dynamics in clean nanowire-SC junctions and may contribute to understanding of the Josephson effect in ballistic topological junctions on semiconductor nanowires.

Here we report observation of half-integer Shapiro steps in an InAs nanowire Josephson junction.
We observed the half-integer voltage steps at $n/2 \cdot hf/2e~ (n=1,2,3, ...)$.
From the temperature dependence of the switching current, we confirmed that the CPR of the junction is skewed at low temperatures.
Our numerical calculation for the short ballistic junction CPR reproduces the half-integer steps, and agrees closely with the experiments, including the temperature and gate voltage dependencies of the half-integer steps. 

A Josephson junction is fabricated on a self-assembled InAs single nanowire, which is placed on a Si substrate.
 These nanowires grown by Chemical Beam Epitaxy have 80-nm diameter~\cite{baba2017gate,baba2018cooper,kamata2018anomalous,ueda2019dominant}.
A scanning electron microscopy (SEM) image of the fabricated device is shown in Fig. 1(a).
We note that the device used in the measurement is similar to but different from that in the figure because we were concerned about any possible damage to our experimental device caused by the SEM observation. 
We evaporated 60 nm-thick aluminum (Al) (shown in blue) on the nanowire for making the superconducting contacts with the junction length between the two SCs of approximately 100 nm.
Then, the top gate electrode (orange) of Ti/Au (5 nm/150 nm) was fabricated, following atomic layer deposition of 20-nm-thick $\rm Al_2O_3$. 

We first performed a DC measurement on the device. The voltage $V$ as a function of the bias current $I$ at a top gate voltage of $V_{\rm g} = 0 \rm \ V$ and temperature of $T = 50 \rm \ mK$ is shown in Fig. 1(b). 
The red (blue) curve was obtained when $I$ was swept in the downward (upward) direction.
The supercurrent flows through the nanowire at $V_{\rm g} = 0 \rm \ V$.
The switching current $I_{\rm sw}$ and the retrapping current $I_{\rm r}$ are 40 nA and 38 nA, respectively. 
The small difference between $I_{\rm sw}$ and $I_{\rm r}$ is attributed to thermal heating~\cite{likharev1979superconducting, Courtois2008}.
\begin{figure}[t]
  \centering
    \includegraphics[width=1.0\linewidth]{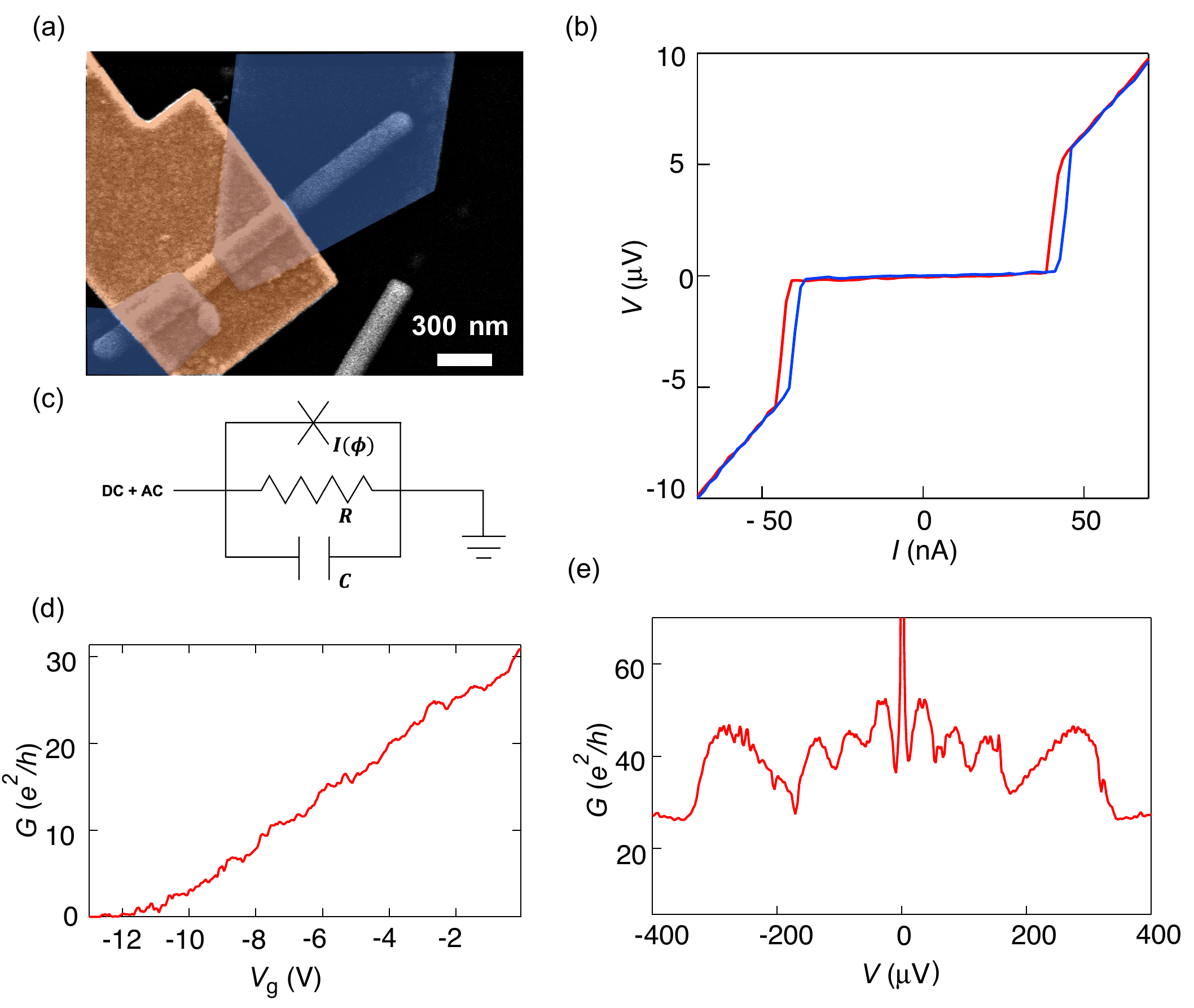}
  \label{Fig.1}
  \caption{
    (a) A SEM image of an InAs nanowire Josephson junction with the top gate electrode (orange region). 
    This device is different from the one used in this study. 
    The junction length between two Als (blue region) is approximately 100 nm. 
    (b) $I$-$V$ curves for our Josephson junction at $V_{\rm g}=\rm 0 \ V$ and $T=\rm50\  mK$. 
    The red (blue) line represents the downward (upward) current sweep. 
    (c) Diagram of an RCSJ circuit. The DC and AC current are applied to the Josephson junction in parallel with the resistance and capacitance. 
    (d) $G$ against $V_{\rm g}$ at $V=\rm 1 \ \mu V$ and $T=\rm 4 \ K$. 
    The pinch-off voltage is $V_{\rm g}=\rm  -12\  V$. 
    (e) $G$ vs. $V$ at $V_{\rm g}=\rm 0 \ V$ and $T=\rm50\  mK$. Multiple Andreev reflection is observed.
  }   
\end{figure}

For the quantitative analysis, we introduced a resistively-capacitively shunted junction (RCSJ) model~\cite{stewart1968current,mccumber1968effect}. 
The equivalent circuit for this model is shown in Fig. 1(c); we used it to study the dynamics of the phase difference between two SCs by solving 
\begin{equation}
  \frac{\hbar C}{2e} \frac{d^2\phi}{dt^2} + \frac{\hbar}{2eR} \frac{d\phi}{dt} + I(\phi) = I_{\rm dc} + I_{\rm ac} \sin(2\pi ft)
\end{equation}
where $C$, $R$, $\phi$, $I(\phi)$, $I_{\rm dc}(I_{\rm ac})$, and $f$ are the junction capacitance, resistance, phase difference between two SCs, the CPR, the applied DC (AC) current, and the applied frequency, respectively. 
This equation is transformed into a dimensionless equation with $t^{'}=(2eI_{\rm sw} R)/\hbar \cdot t$ described as
\begin{equation}
  \beta \frac{d^2\phi}{dt^{'2}} + \frac{d\phi}{dt^{'}} + i(\phi) = i_{\rm dc} + i_{\rm ac} \sin(2\pi f^{'}t^{'})
\end{equation}
Here we define $\beta = 2eI_{\rm sw} R^2 C/\hbar$ as the Stewart-McCumber parameter~\cite{stewart1968current,mccumber1968effect} and $f^{'}=f/2eI_{\rm sw} R$.
The $\beta$ value of 0.008 was estimated from the geometry of the junction, with $C=130 \rm \ fF$, $R= 800 \rm \ \Omega$ at $V_{\rm g}=0 \rm \ V$, and $I_{\rm sw}=40.7 \rm \ nA$.
The small value of $\beta$ means that the junction is highly overdamped.
 
The differential conductance $G$ as a function of $V_{\rm g}$ is shown in Fig. 1(d) for $T = 4 \rm \ K$, which is larger than the Al critical temperature. 
The pinch-off voltage depleting the carrier density of the nanowire is $V_{\rm g} = -12 \rm \ V$. 

We investigated $G$ vs. $V$ at $V_{\rm g}=0 \rm \ V$ and $T=\rm50\  mK$ as shown in Fig. 1(e). 
We observed several conductance peaks attributed to the multiple Andreev reflection. 
Peaks are expected to appear at $V= 2\Delta/en~ (n=1,2,3, ...)$ with $\Delta=140 \rm \ \mu eV$ of the Al superconducting gap energy. 
Indeed, the observed peaks are located at these voltages for $n$ = 1 to 4. 
Observation of the multiple Andreev reflection up to the fourth order implies that elastic Andreev reflections occur sequentially at both interfaces of the Al and the nanowire. 
Therefore, we assumed that the interfaces are transparent enough and the nanowire between the interfaces is clean enough to enable ballistic transport through the Al-nanowire-Al junction at $V_{\rm g}=0 \rm \ V$~\cite{hurd1996current}.

Next, we applied microwave radiation to study the Shapiro response.
The black curve in Fig. 2(a) represents the $I$-$V$ curve at $V_{\rm g}=0 \rm \ V$ and $T=\rm50\  mK$ measured for a microwave input with applied power $P = 6 \rm \ dBm$ and $f = 4.2 \rm \ GHz$. 
The conventional Shapiro steps are observed at $V= nhf/2e~ (n=1,2,3, ...)$.
Furthermore, there are additional plateaus at the half-integer multiples of $hf/2e$.
The half-integer steps are detected at various microwave frequencies (see the supplementary material).
$R$ as a function of $I$ becomes zero at the integer steps as shown by the red dotted line in Fig 2(a).
The half-integer steps appear as the dips.

To determine the cause of the half-integer steps, we introduced CPR for a short ballistic Josephson junction having a single channel which is given by 
\begin{equation}
  \begin{split}
  I(\phi) = \frac{e\Delta(T)}{2\hbar} \frac{\tau \sin (\phi)}{[1-\tau \sin^2(\phi/2)]^{1/2}}\\
   \times \tanh \left(\frac{\Delta(T)}{2k_B T}[1-\tau \sin^2(\phi/2)]^{1/2}\right)
  \end{split}
  \label{Eq. (3)}
\end{equation}
where $\Delta$, $\tau$, and $T$ are the superconducting gap, transmission of the junction, and temperature, respectively~\cite{kulik1978josephson,haberkorn1978theoretical,beenakker1992three}. 
This CPR is skewed from the sinusoidal function of $\phi$ for the conventional Josephson junctions. 
The CPR with $\tau=0.98$ and $T= 50 \rm \ mK$ is represented by the red line in Fig. 3(a) as the normalized supercurrent vs. $\phi$. 
As previously predicted, the skewed CPR can generate anomalous steps at the fractional quantized voltages because the skewed CPR includes higher harmonic components whose periodicity is fractions of $2\pi$.
\begin{figure}[]
  \centering
    \includegraphics[width=1.0\linewidth]{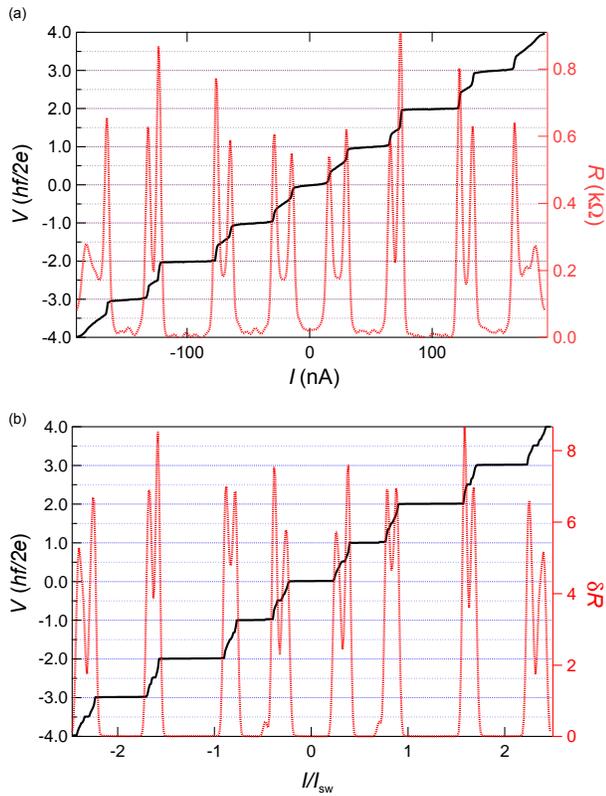}
  \label{Fig.2}
  \caption{
    (a) Shapiro steps under microwave irradiation at $f=\rm 4.2 \ GHz$, $V_{\rm g}=\rm 0\  V$, and $P = 6 \rm \ dBm$ (black curve). 
    Both conventional Shapiro steps and half-integer steps are observed. 
    The red curve represents $R$ of the $I$-$V$curve. 
    (b) Numerical simulated Shapiro response at $\beta =0.008$, $f^{'}=0.095$, and $i_{\rm ac}=2$. The half-integer steps are reproduced.
  }   
\end{figure}

We numerically calculated the Shapiro responses using the CPR with $\tau=0.98$ and $T= 50$ mK in the RCSJ model. 
In Fig. 2(b) a typical result is shown of the calculated Shapiro steps for $f = 4.2 \rm \ GHz$, $\beta = 0.008$, and $I_{\rm sw}=40 \rm \ nA$.
The black solid and red dotted lines represent $V$ vs. $I/I_{\rm sw}$ and the differential resistance $\delta R$ vs. $I/I_{\rm sw}$, respectively. 
It is clear that the experimentally observed half-integer Shapiro steps are reproduced by the numerical calculation.
In addition, we recognize weak features of additional steps with height of one-third and two-thirds of the quantized voltages in the calculation. 
However, in the experiment we observed neither of them (see Fig.2(a)), but the half-integer plateaus were largely tilted.
We therefore suspect that the one-third and two-third steps were smeared in the experiment because of the insufficient resolution of our measurement setup.
In our numerical calculation, we assumed all the channels have the same $\tau$ for simplicity. 
Even with this rough assumption, the experimental results are well explained by the CPR skewness as shown in the following discussion.
 
The CPR skewness depends on $\tau$ and $T$. Therefore, we investigated the $V_{\rm g}$ and $T$ dependence of $I_{\rm sw}$.
From the temperature dependence of $I_{\rm sw}$, we can evaluate $\tau$ because $I_{\rm sw}$ is derived as the maximum of the CPR in Eq. (3).
Then, we measured the temperature dependence of $I_{\rm sw}$ at $V_{\rm g}= \rm \ 0, -1, -2,$ and $-3$ V as shown by the red, yellow, blue, and purple circles in Fig. 3(b), respectively. 
We carried out the numerical fitting of the experimental results with the maximum of Eq. (3) with $\tau$ and effective channel number as free fitting parameters. 
The calculated result is shown as the solid lines in Fig. 3(b).
We obtained excellent fitting of the solid lines to the experimental results for $V_{\rm g}= \rm \ 0, -1, -2, and -3$ V with $\tau$ of 0.98, 0.85, 0.89, and 0.7, respectively, as shown in Fig. 3(a).
In particular, $\tau=0.98$ for $V_{\rm g} = 0 \rm \ V$ is nearly unity, indicating nearly perfect transmission. 
The CPR or normalized supercurrent vs. $\phi$ with $\tau= \rm \ 0.98, 0.85, 0.89, 0.7$ at $T= 50 \rm \ mK$ is represented in Fig. 3(a). 
The results indicate that the CPR skewness is remarkable at $V_{\rm g}= 0 \rm \ V$ because of the high transmission.
 \begin{figure}[b]
  \centering
    \includegraphics[width=1.0\linewidth]{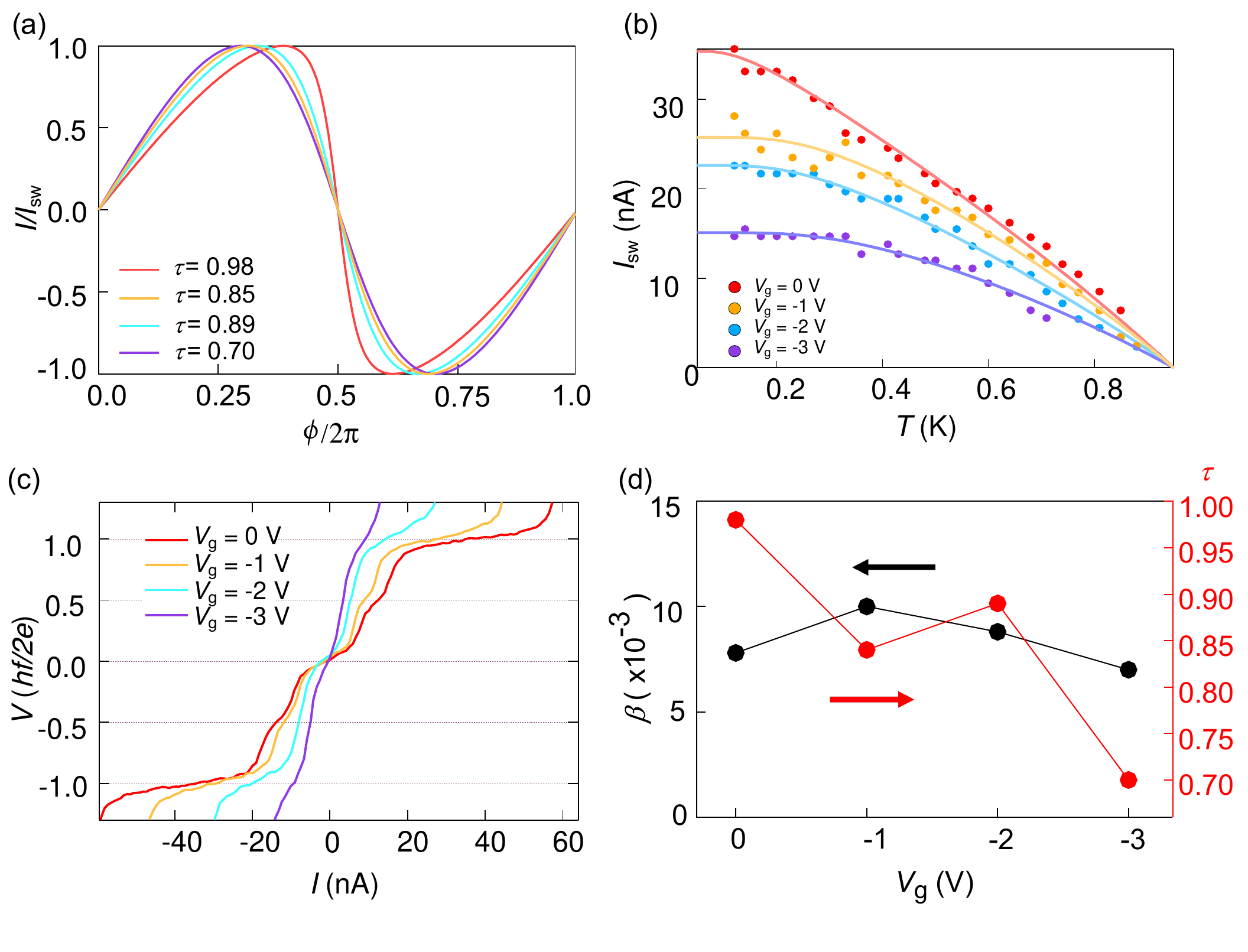}
  \label{Fig.3}
  \caption{
    (a) Comparison of short ballistic CPR curves for $\tau=0.98$ (red), $\tau=0.85$ (orange), $\tau=0.89$ (blue), and $\tau=0.7$ (purple) at $T = 50$ mK. 
    (b) $I_{\rm sw}$ vs. $T$. The red, orange, blue, and purple dots correspond to $V_{\rm g}=\rm 0, -1, -2, -3\ V$, respectively. 
    These data were fit using the short ballistic CPR (see the supplementary material). (c) $I$-$V$ curves at $V_{\rm g}=\rm 0, -1, -2,$ and $-3$ V with $P=\rm \ -17$ dBm and $f=\rm 1.8$ GHz.
    (d)$\beta$ and $\tau$ vs. $V_{\rm g}$. 
    $\tau$ decreases whereas $\beta$ is nearly constant with decreasing $V_{\rm g}$.
  }   
\end{figure}

The $V_{\rm g}$ dependence of the Shapiro steps measured for $f= 1.8 \rm \ GHz$, $P = -17 \rm \ dBm$ and $T=\rm50\  mK$ is shown in Fig. 3(c).
The half-integer steps vanish before the integer steps do with decreasing $V_{\rm g}$. 
First we checked $\beta$ vs. $V_{\rm g}$ because the junction dynamics in the RCSJ model highly depends on $\beta$.
The obtained $\beta$ vs. $V_{\rm g}$ is shown in Fig. 3(d).
$\beta$ is as small as 0.008 to 0.01 in the $V_{\rm g}$ range of our measurement, indicating that the RCSJ circuit is highly overdamped.
Thus, the measured $V_{\rm g}$ dependence of the half-integer steps cannot be attributed to the change in $\beta$.
On the other hand, as shown in Fig. 3(a), the CPR skewness fades away or $\tau$ decreases as $V_{\rm g}$ decreases.
Therefore, the disappearance of the half-integer steps with decreasing $V_{\rm g}$ is attributed to the change in CPR skewness.
\begin{figure}[t]
  \centering
    \includegraphics[width=1.0\linewidth]{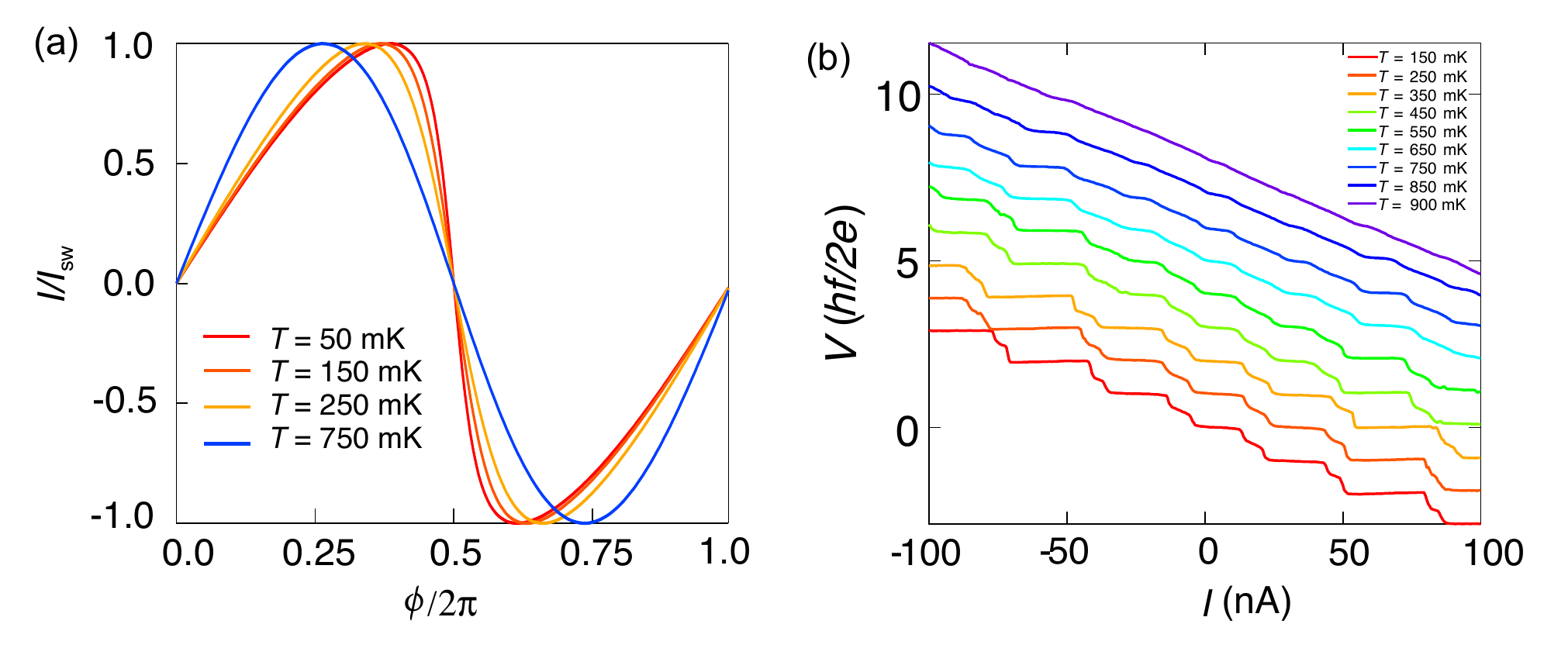}
  \label{Fig.4}
  \caption{
    (a) Comparison of short ballistic CPR curves for $T=\rm 50 \ mK$ (red), $T=\rm 150 \ mK$ (orange), $T=\rm 250 \ mK$ (yellow) and $T=\rm 750 \ mK$ (blue) at $\tau=0.98$. 
    $I_{\rm sw}$ is the simulated critical current at each temperature for $\tau=0.98$. 
    (b) Temperature dependence of the Shapiro steps at $P = \rm 11 \ dBm$ and $f=\rm  4.2 \rm \ GHz$.
    The y axis shows the data for 150 mK; the other data are incremented upward by $hf/2e$. 
  } 
\end{figure}

Finally, to confirm the relation between the CPR skewness on the half-integer steps, we investigated the temperature dependence of the Shapiro steps at $V_{\rm g} = 0 \rm \ V$.
The calculated CPR with $\tau=0.98$ at $T= 50, 150, 250, 750 \rm \ mK$ is shown in Fig. 4(a).
As $T$ increases, the CPR becomes closer to a sinusoidal function and the skewness disappears.
The $I$-$V$ curves including the Shapiro steps measured at various temperatures between $T=150 - 900 \rm \ mK$ are shown in Fig. 4(b).
Both the conventional integer and half-integer steps become gradually vague as $T$ increases.
In addition, the half-integer steps almost vanish at $T=750 \rm \ mK$, whereas the conventional steps are still visible at the even higher $T$.
This behavior is consistent with the temperature dependence expected from the CPR as shown in Fig. 4(a). 
Consequently, we conclude that the observed half-integer steps originate from the skewed CPR.
 
We note that the observed half-integer steps are not related to the $0$-$\pi$ transition of the junction, which also generates half-integer steps as reported in previous research~\cite{Sellier2004, Frolov2006, stoutimore2018second}.
The $0$-$\pi$ transition can be invoked in a Josephson junction with the time-reversal symmetry broken~\cite{Hart2016, Murani2017} or the quantum dot Josephson junctions~\cite{Dam2006, jorgensen2007critical, Eichler2009, li20170}.
We detected the anomalous steps in the time-reversal invariant system, that is, our junction does not contain ferromagnetic materials and we applied no magnetic field.
In addition, quantum dots are not formed in our nanowire as confirmed from the conductance as a function of $V_{\rm g}$.
Therefore, we can rule out these scenarios.
 
In conclusion, we observed half-integer Shapiro steps in the short ballistic Josephson junction of an InAs nanowire.
We associated this observation with the skewed CPR in a short ballistic Josephson junction by using numerical calculations.
This association was further supported by the experimentally derived junction transmission from the temperature dependence of the switching current.
The present study elucidated the relation between the junction CPR and the junction dynamics of the AC Josephson effect, and may provide appropriate knowledge and  methods for exploring the topological phenomena of the proximitized superconducting dynamics that is realized in superconductor-nanowire junctions.

We thank Prof. S. Jeppesen for the material growth.
This work was partially supported by 
a Grant-in-Aid for Scientific Research (B) (grant no. JP18H01813)
; a Grant-in-Aid for Scientific Research (S) (grant nos. JP19H05610)
; JSPS Program for Leading Graduate Schools (MERIT)
; JSPS Research Fellowship for Young Scientists (grant nos. JP19J13867, JP18J14172) from JSPS
; JST PRESTO (grant no. JPMJPR18L8)
; the Ministry of Science and Technology of China (MOST) through the National Key Research and Development Program of China (grant nos. 2016YFA0300601 and 2017YFA0303304)
; the National Natural Science Foundation of China (grant nos. 91221202, 91421303, and 11874071)
; and the Swedish Research Council (VR).

\bibliographystyle{apsrev4-1}
%

\end{document}